# Characterizing the Habitable Zones of Exoplanetary Systems with a Large Ultraviolet/Visible/Near-IR Space Observatory

Kevin France[1*], Evgenya Shkolnik[2], Jeffrey Linsky[1], Aki Roberge[3], Thomas Ayres[1], Travis Barman[4], Alexander Brown[1], James Davenport[5], Jean-Michel Desert[1], Shawn Domagal-Goldman[3], Brian Fleming[1], Juan Fontenla[6], Luca Fossati[7], Cynthia Froning[8], Gregg Hallinan[9], Suzanne Hawley[5], Renyu Hu[10], Lisa Kaltenegger[11], James Kasting[12], Adam Kowlaski[13], Parke Loyd[1], Pablo Mauas[14], Yamila Miguel[15], Rachel Osten[16], Seth Redfield[17], Sarah Rugheimer[18], Christian Schneider[19], Antigona Segura[20], John Stocke[1], Feng Tian[21], Jason Tumlinson[16], Mariela Vieytes[14], Lucianne Walkowicz[22], Brian Wood[23], and Allison Youngblood[1]

[1]University of Colorado, [2]Lowell Observatory, [3]NASA/GSFC, [4]U. Arizona, [5]U. Washington, [6]NWRA, [7]U. Bonn, [8]U. Texas, [9]Caltech, [10]JPL, [11]Cornell, [12]Penn State, [13]U. Maryland, [14]IAFE, [15]Observatoire de la Côte d'Azur, [16]STScI, [17]Wesleyan, [18]Harvard, [19]ESA/ESTEC, [20]UNAM, [21]Tsinghua U., [22]Adler Planetarium, [23]NRL

Understanding the surface and atmospheric conditions of Earth-size ($R_P \approx 1 - 1.5$ $R_\oplus$), rocky planets in the habitable zones (HZs) of low-mass stars is currently one of the greatest astronomical endeavors. The Astro2010 Decadal Survey ranks the **"Identification and characterization of nearby habitable exoplanets"** as the premier Discovery Goal for astrophysics at present, with **"can we identify the telltale signs of life on an exoplanet?"** as a specific focus question. The nearest known Super-Earth mass planets ($2 - 10$ $M_\oplus$) in the HZ orbit late-type stars (M and K dwarfs). In addition, all of the HZ planets found by *TESS* will be around M dwarfs, making these systems prime targets for spectroscopic biomarker searches with *JWST* and future direct spectral imaging missions (Deming et al. 2009; Belu et al. 2011).

The planetary effective surface temperature alone is insufficient to accurately interpret biosignature gases when they are observed in the coming decades. The UV stellar spectrum drives and regulates the upper atmospheric heating and chemistry on Earth-like planets, is critical to the definition and interpretation of biosignature gases (e.g., Seager et al. 2013), and may even produce false-positives in our search for biologic activity (Hu et al. 2012; Tian et al. 2014; Domagal-Goldman et al. 2014). Therefore, our quest to observe and characterize biological signatures on rocky planets *must* consider the star-planet system as a whole, including the interaction between the stellar irradiance and the exoplanetary atmosphere.

Spectral observations of $O_2$, $O_3$, $CH_4$, and $CO_2$, are expected to be important signatures of biological activity on planets with Earth-like atmospheres (Des Marais et al. 2002; Kaltenegger et al. 2007; Seager et al. 2009). The chemistry of these molecules in the atmosphere of an Earth-like planet depends sensitively on the strength and shape of the host star's UV spectrum. $H_2O$, $CH_4$, and $CO_2$ are sensitive to far-UV radiation (FUV; $100 - 175$ nm), in particular the bright HI Lyα line, while the atmospheric oxygen chemistry is driven by a combination of FUV and near-UV (NUV; $175 - 320$ nm) radiation (Fig 1). Furthermore, the FUV and NUV bandpasses themselves may be promising spectral regimes for biomarker detection (Bétrémieux & Kaltenegger 2013). High levels of extreme-UV (EUV; $10 - 91$ nm) irradiation can drive hydrodynamic mass loss on all types of exoplanets (e.g., Koskinen et al. 2013; Lammer et al. 2014). Ionization by EUV photons and the subsequent loss of atmospheric ions to stellar wind pick-up can also drive wide-scale atmospheric mass loss on geologic time scales. As the opacity of the ISM precludes direct observations of the EUV irradiance, FUV observations provide the best constraints on the EUV emission from cool stars (Linsky et al., 2014; Shkolnik & Barman 2014).

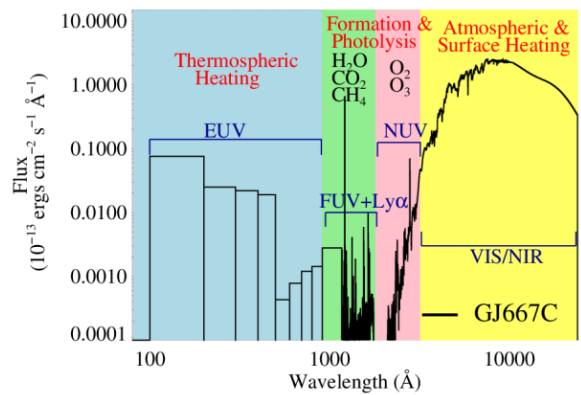

**Figure 1**: *HST spectrum of GJ 667C, illustrating the influence of each spectral bandpass on the atmosphere of an Earth-like planet orbiting this star (France et al. 2013, Linsky et al. 2014). GJ 667C hosts a super-Earth mass planet located in the HZ.*

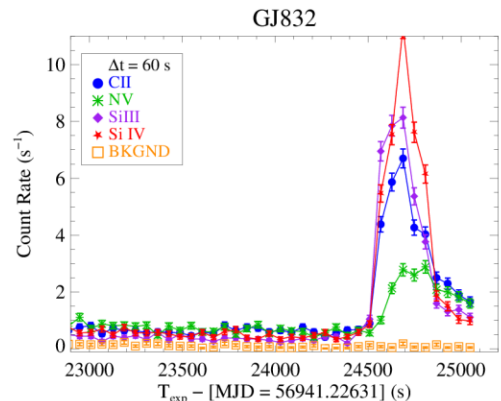

*Figure 2*: *FUV flare on the "optically inactive" M dwarf GJ 832 (d = 4.9 pc), illustrating a flare event in FUV emission lines on sub-minute timescales (HST-COS data, binned to 60 s intervals).*

The temporal variability of these sources must also be considered as even weakly-active, "old" (several Gyr) M dwarfs show extreme flare events in their UV light curves (factors of greater than 10 flux increases on time scales of minutes, see Fig. 2; France et al. 2013; Loyd & France 2014). Impulsive UV events are also signposts for energetic flares, those that are associated with large ejections of charged particles. Energetic particle deposition into the atmosphere of an Earth-like planet during a large M dwarf flare can lead to significant atmospheric $O_3$ depletions (> 90% for large flares; Segura et al. 2010). This alters the atmospheric chemistry and increases the penetration depth of UV photons that could potentially sterilize (or catalyze) surface life. Given that particle fluxes cannot typically be directly measured for stars other than the Sun, UV observations offer the best estimates of these important particle environments.

NASA's next large ultraviolet/optical/infrared flagship mission, a LUVOIR Surveyor, could carry out both the direct detection of atmospheric tracers on these worlds and the essential characterization of the star-planet system. The transformative science enabled by a LUVOIR Surveyor to the field of astronomy as a whole, including exoplanet (ExoPAG) and cosmic origins (COPAG) themes, is made clear in NASA's 2013 Astrophysics Roadmap (*Enduring Quests, Daring Visions: NASA Astrophysics in the Next Three Decades*): a LUVOIR Surveyor (or ATLAST/HDST-like mission; Tumlinson et al. 2015) directly addresses the largest number of core NASA science priorities in a single mission in the next 30 years, working toward six "primary goals", *twice as many core science goals as any other intermediate-term mission concept* (Section 6.4 in 2013 Roadmap).

There are a set of baseline requirements for the characterization of the energetic radiation environments around low-mass exoplanet host stars. Spectral coverage to wavelengths as short as 102 nm is crucial to characterizing the star in important activity tracers (e.g., O VI and Lyα), as well as peak absorption cross-sections for relevant molecular species in planetary atmospheres (e.g., $O_2$ and $H_2$). We would like to characterize typical mid-M dwarf exoplanet host stars out to 30 pc with sufficient signal-to-noise to carry out time-resolved, $R \approx 3000$ spectroscopy of UV emission lines (e.g., N V, Si IV, C IV; $F_\lambda \sim 1 \times 10^{-16}$ erg cm$^{-2}$ s$^{-1}$) on the scale of minutes (S/N ~ 10 per 60 s). These sensitivities can be achieved with a 10-meter class mirror, optical coatings with broadband UV reflectivity ≥ 90%, and photon-counting detectors with UV DQE ≥ 75%. These technology goals can be achieved with a reasonable investment in laboratory development efforts and hardware demonstration flights (via suborbital and small satellite missions; e.g., France et al. 2012) in the next 5 – 10 years, enabling the start of a LUVOIR Surveyor mission in the 2020s.